\documentclass[a4paper,11pt]{article}
\usepackage{pos}

\title{Discovery Forecasts of the Diffuse Ultra-High-Energy Neutrino Flux with IceCube-Gen2}
 \ShortTitle{UHE neutrino flux discovery forecast}

\author*[a]{Victor B. Valera}
\author[a]{Mauricio Bustamante}
\author[b]{Christian Glaser}

\affiliation[a]{Niels Bohr International Academy, Niels Bohr Institute, University of Copenhagen, \\ DK-2100 Copenhagen, Denmark}

\affiliation[b]{Department of Physics and Astronomy, Uppsala University,\\ Uppsala, SE-752 37, Sweden}

\emailAdd{vvalera@nbi.ku.dk}
\emailAdd{mbustamante@nbi.ku.dk}
\emailAdd{christian.glaser@physics.uu.se}

\abstract{The discovery of ultra-high-energy (UHE) neutrinos has the potential to offer unique insight into fundamental questions. To capitalize on the upcoming opportunity provided by new UHE neutrino telescopes, we provide state-of-the-art forecasts of the discovery of a diffuse flux of UHE neutrinos over the next 10-20 years, focusing on neutrino radio-detection in the planned IceCube-Gen2 detector. We use state-of-the-art flux predictions and detector modeling. We find that, even under conservative analysis choices, most benchmark UHE neutrino flux models from the literature may be discovered within 10 years of detector exposure, with many discoverable sooner, and may be distinguished from each other. Our results demonstrate the transformative potential of next-generation UHE neutrino telescopes.}

\ConferenceLogo{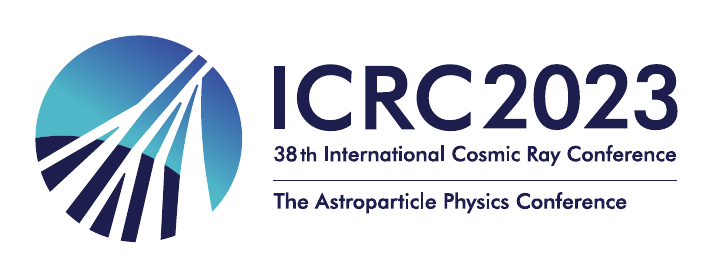}

\FullConference{%
38th International Cosmic Ray Conference (ICRC2023)\\
  26 July - 3 August, 2023\\
  Nagoya, Japan}

\begin{document}
\maketitle

\section{Introduction}
Ultra-high-energy (UHE) neutrinos with energies in the EeV scale (1~EeV = $10^{18}$ eV) were predicted in the late 1960s~\cite{Berezinsky:1969erk} as a result of the interaction between UHE cosmic rays and cosmological photon fields~\cite{Greisen:1966jv, Zatsepin:1966}. They offer valuable insights into astrophysics and particle physics at the highest energies. UHE neutrinos remain undetected, motivating the construction of larger neutrino telescopes and the consideration of various flux predictions~\cite{Ackermann:2022rqc}.

IceCube-Gen2 is among the largest futre UHE neutrino telescopes in planning. It aims to discover UHE neutrinos by detecting the radio signals that they induce upon interacting in the Antarctic ice~\cite{IceCube-Gen2:2020qha}. In this study, we provide detailed forecasts to assess the discovery potential of UHE neutrinos using IceCube-Gen2. This proceeding is based on Ref.~\cite{Valera:2022wmu}, where a detailed description of our methods and results can be found. We consider a wide range of benchmark flux models~\cite{IceCube:2020wum, IceCube:2021uhz, Heinze:2019jou, Fang:2013vla, Padovani:2015mba, Fang:2017zjf, Muzio:2019leu, Rodrigues:2020pli, Anker:2020lre, Muzio:2021zud}. The calculations incorporate state-of-the-art ingredients at every stage, including flux models, neutrino-nucleon cross section~\cite{Bertone:2018dse}, neutrino propagation through the Earth~\cite{Garcia:2020jwr}, neutrino detection~\cite{Glaser:2019cws, Glaser:2019rxw}, and backgrounds~\cite{Fedynitch:2018cbl}.

Our results show promising prospects for the discovery of UHE neutrinos. Most benchmark flux models are expected to be discovered within a few years, even under conservative analysis choices. Less conservative choices yield even better prospects. Our goal is to provide a realistic assessment of the science potential of upcoming UHE neutrino telescopes, taking into account experimental and theoretical nuances that are often overlooked. Our methods can be adapted for other telescopes, and facilitate the assessment and comparison of competing designs.

\section{Ultra-high-energy neutrinos}
\begin{figure*}[t!]
 \centering
 \includegraphics[width=\textwidth]{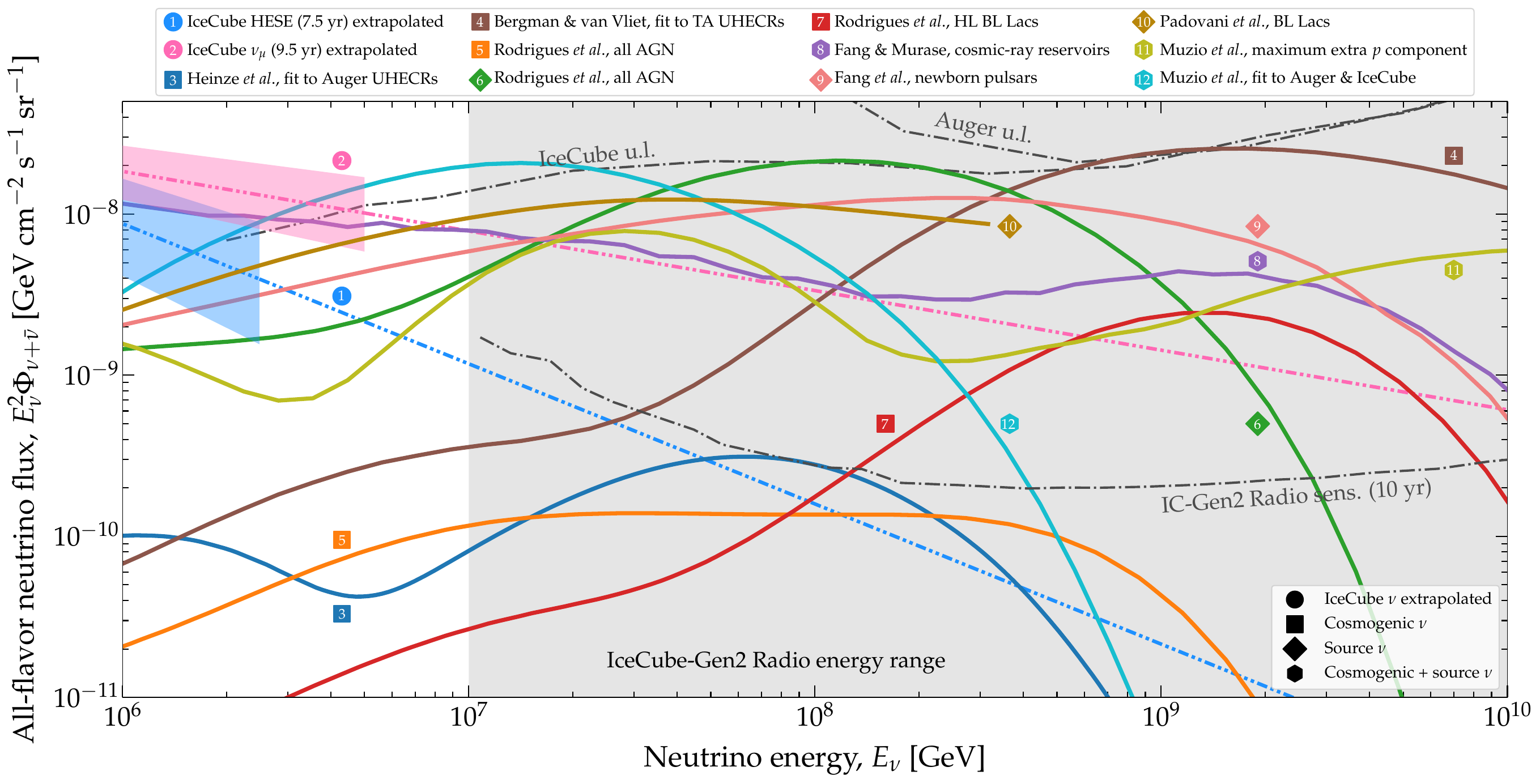}
 \caption{\label{fig:benchmark_spectra} Benchmark diffuse ultra-high-energy neutrino flux models~\cite{Fang:2013vla, Padovani:2015mba, Fang:2017zjf, Heinze:2019jou, Muzio:2019leu, Rodrigues:2020pli, Anker:2020lre, IceCube:2020wum,  Muzio:2021zud, IceCube:2021uhz} used here to assess the flux discovery capabilities of the radio array of IceCube-Gen2~\cite{IceCube-Gen2:2020qha} (``IceCube-Gen2 Radio").  The upper limits on the flux are from IceCube~\cite{IceCube:2018fhm} and the Pierre Auger Observatory~\cite{PierreAuger:2019ens}. Figure from Ref.~\cite{Valera:2022wmu}.}\label{fig:fluxes}
\end{figure*}
Ultra-high-energy neutrinos can be classified as \textit{source neutrinos} if they are produced within the UHECR sources, or \textit{cosmogenic neutrinos} if they are produced during their journey to Earth. The production mechanism involves the decay of charged pions produced in the interactions of UHECR protons with matter or radiation. Each final-state neutrino carries approximately $5\%$ of the energy of the parent proton. The energy spectra of UHE neutrinos depend on the production mechanism. Neutrinos produced in $pp$ interactions follow a power-law spectrum similar to that of the parent protons. Neutrinos produced in $p\gamma$ interactions have energy spectra determined by the spectra of the parent protons and photons, with a characteristic energy set by the requirements to produce a $\Delta$ resonance.

Figure~\ref{fig:fluxes} shows the energy spectra of benchmark UHE neutrino flux models used in the study~\cite{Valera:2022wmu}. These models represent a wide range of predictions, from optimistic to pessimistic, reflecting the uncertainties in UHECR properties and their sources. The models used in this study are the same as those used in a previous work, and their detailed features can be found in the respective references~\cite{Valera:2022wmu, Valera:2022ylt}.

\section{Neutrino propagation inside the Earth}
Upon reaching the surface of the Earth, UHE neutrinos propagate underground toward the detector. Interactions with matter underground play a significant role in attenuating the neutrino flux that reaches the detector. The attenuation depends on the energy, direction, and flavor of the neutrinos. The in-Earth propagation of UHE neutrinos is accounted for in detail in the forecasts, considering the journey from the surface to the IceCube-Gen2 radio array. At energies above a few GeV, the primary interaction channel for neutrinos is neutrino-nucleon deep inelastic scattering (DIS). The state-of-the-art BGR18 calculation of the cross sections~\cite{Bertone:2018dse} is adopted for the propagation and detection of neutrinos.

The severity of in-Earth propagation effects on the neutrino flux depends on the neutrino energy and direction, measured by the zenith angle $\theta_z$. Higher energies and longer path lengths lead to stronger effects. Upgoing neutrinos are more heavily attenuated, and virtually no upgoing neutrinos reach the detector for UHE neutrinos unless the surface flux is exceptionally large. Therefore, the forecasts primarily consider downgoing and horizontal neutrinos.

For a detailed calculation of the Earth effect on the neutrino flux, we employ the Monte Carlo code {\sc NuPropEarth}~\cite{Garcia:2020jwr}. This code accounts for the dominant contributions from CC and NC neutrino-nucleon DIS, as well as the subdominant effects of other interaction channels and regeneration processes. The internal matter density profile of the Earth is based on the Preliminary Reference Earth Model.

\section{Neutrino event rate computation}
To predict the event rates of neutrino-induced signals in the radio array of IceCube-Gen2, we employ the methods outlined in Refs.~\cite{Valera:2022ylt, Valera:2022wmu}. The details of these methods can be found in Ref.~\cite{Valera:2022wmu}, while we provide a brief overview here.

In-ice, radio-based neutrino telescopes measure the radio signals in the ice emitted by particle showers through the Askaryan effect~\cite{Askaryan:1961pfb}. In a shower resulting from a neutrino-nucleon ($\nu N$) deep inelastic scattering (DIS) event, the shower energy ($E_{\rm sh}$) is a fraction of the parent neutrino energy ($E_\nu$). The specific fraction depends on the flavor of the interacting neutrino and whether the interaction is neutral current (NC) or charged current (CC). In the detector, after a $\nu N$ DIS event, Askaryan radiation propagates through the ice, attenuating en route to the detector. The radiation may or may not trigger the antennas upon reaching the detector, depending on various factors such as the shower energy, shower direction, antenna characteristics, and the size and geometry of the detector array. We account for these factors using dedicated Monte Carlo simulations of neutrino-induced shower production, propagation, and detection, employing the state-of-the-art tools NuRadioMC~\cite{Glaser:2019cws} and NuRadioReco~\cite{Glaser:2019rxw}, which are also used by the IceCube-Gen2 Collaboration. These simulations characterize the expected detector response and are described by the detector effective volume, $V_{\rm eff}$, which depends on the shower energy, direction, and interaction type.

To account for the limited energy and angular resolution of the detector, we introduce energy and angular resolution functions and express the event rate in terms of reconstructed shower energy $E_{\rm sh}^{\rm rec}$ and reconstructed direction $\theta_z^{\rm rec}$. The energy resolution function is modeled as a Gaussian in $\log_{10} E_{\rm sh}$ centered at $\log_{10} E_{\rm sh}^{\rm rec}$ and with a width of $\sigma_\epsilon = 0.1$, corresponding to $10\%$ of an energy decade. The angular resolution function is modeled as a Gaussian centered $\theta_z = \theta_z^{\rm rec}$ and with a width of $\sigma_\theta = 2^\circ$.

\section{Backgrounds}
We discuss the potential to discover and differentiate between benchmark UHE neutrino flux models 1--12 from Fig.~\ref{fig:fluxes}. We take into account two main sources of background: atmospheric muons and the UHE tail of the IceCube high-energy neutrino flux.

\textit{Atmospheric muons:} Regarding atmospheric muons, they can produce in-ice showers that generate a small, but unavoidable background for UHE neutrino searches. The rate of muon-induced events in the radio array of IceCube-Gen2 is estimated using the hadronic interaction model {\sc Sybill 2.3c} and applying a surface veto. The resulting energy and angular distribution of muon-induced events shows a concentration at low energies ($E_{\rm sh}^{\rm rec} \lesssim 10^8$ GeV) and in downward directions ($\cos \theta_z^{\rm rec} \gtrsim 0$). The all-sky integrated rate of muon-induced events above $10^8$ GeV, which cannot be vetoed by the surface veto, is lower than 0.1 event per year. Therefore, atmospheric muons represent an obstacle mainly to the discovery of a small UHE neutrino flux that peaks at low neutrino energies.

\textit{The UHE tail of the IceCube high-energy neutrinos:} The current measurements by IceCube cover the energy range from 10 TeV to a few PeV, with sparse data above the PeV range due to the steeply falling neutrino energy spectrum. It is presently unknown if the measured flux extends to ultra-high energies (beyond 100 PeV) and what its spectrum looks like at those energies. Benchmark flux models 1 and 2 are extrapolations of the IceCube TeV--PeV power-law flux measurements, showing that if the UHE tail of the IceCube flux is significant enough to trigger events in the radio array of IceCube-Gen2, it would contribute as a background to the discovery of UHE neutrino flux models 3--12. The detection of the UHE tail of the IceCube high-energy neutrino flux depends on the spectral index and the presence of a further suppression at or above the few-PeV scale. The spectral index value varies depending on the set of IceCube events used for the fit. A harder spectrum implies a more prominent UHE tail, which is more likely to trigger events in the radio array. 

\section{Flux discovery}
\begin{figure}[t!]
 \centering
 \includegraphics[width=0.5\textwidth]{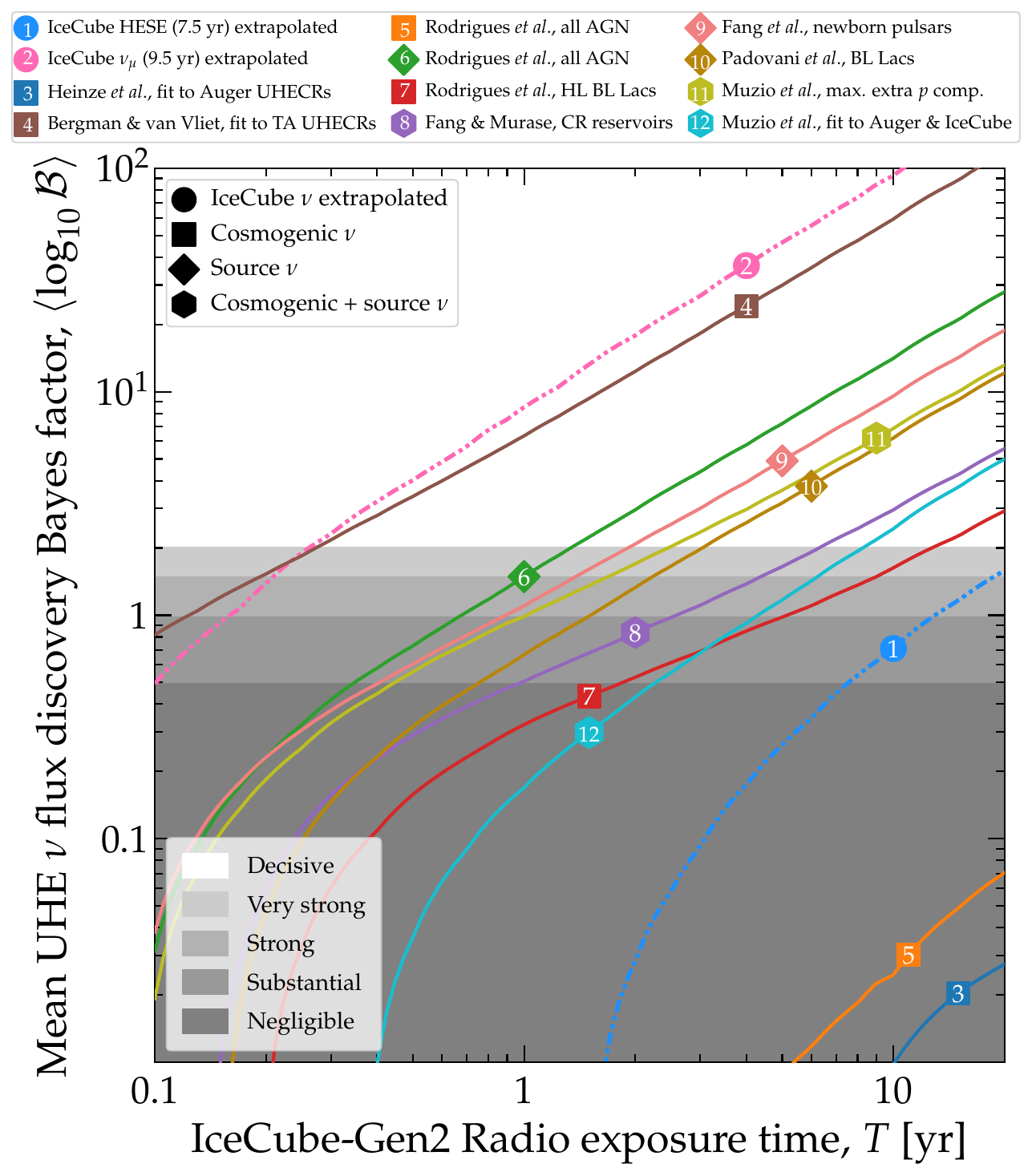}
 \caption{\label{fig:bayes_factor_hard}Discovery potential of benchmark diffuse UHE neutrino flux models 1--12~\cite{Fang:2013vla, Padovani:2015mba, Fang:2017zjf, Heinze:2019jou, Muzio:2019leu, Rodrigues:2020pli, Anker:2020lre, IceCube:2020wum,  Muzio:2021zud, IceCube:2021uhz} in IceCube-Gen2.  The background to discovery consists of atmospheric muons, for all models, plus the tentative UHE tail of the IceCube 9.5-year through-going $\nu_\mu$ flux~\cite{IceCube:2021uhz}, for models 3--12. Figure from Ref.~\cite{Valera:2022wmu}}
\end{figure}

The discovery potential is quantified using a binned likelihood function that compares the predicted and observed event rates in reconstructed shower energy and direction bins. The likelihoods for the signal and background-only hypotheses are computed, and the posterior probability distributions and evidence are obtained using Bayesian methods.

\subsection{Statistical model}
For a given choice of UHE neutrino flux model, $\mathcal{M}_{\rm UHE}$, out of models 3--12 in Fig.~\ref{fig:benchmark_spectra}, and for a given choice of the background UHE tail of the IceCube high-energy neutrino flux, $\mathcal{M}_{\rm HE}$, we quantify the discovery potential on the basis of a Poisson likelihood function binned in reconstructed shower energy and direction, Both for the signal hypothesis (s+bg), and for the background hypothesis (bg). The model parameters are $\boldsymbol\theta \equiv (\log_{10} f_\sigma, \log_{10} (E_{\nu, {\rm cut}}^{\rm HE} / {\rm GeV}))$, and they represent the free parameters on which the neutrino-induced event rate depends: the $\nu N$ cross section, $f_\sigma \equiv \sigma_{\nu N}/\sigma_{\nu N}^{\rm std}$, and the cut-off energy of the background IceCube high-energy neutrino flux, $E_{\nu, {\rm cut}}^{\rm HE}$. In particular, when testing the discoverability of models 1 and 2, we only consider as background the atmospheric muons.

In the Poisson likelihood function, the number of observed events, $N_{{\rm obs}}$, is obtained as a realization of the true observed spectrum, obtained as a Poisson fluctuation for the predicted rate for $\boldsymbol{\theta} = (0, \infty)$. On the other hand the predicted number of events, $N_{{\rm pred}}$, is obtained directly from the prediction of the event rate for a choice of $\boldsymbol{\theta}$. The Bayesian evidence for the signal and background hypothesis is obtained integrating the likelihood over the parameter space of $\boldsymbol{\theta}$, multiplied by the prior distribution. We use wide flat priors for the model parameters $\boldsymbol{\theta}$. Finally the Bayes factor is obtained as a the ratio between the Bayesian evidence for the signal and the background hypothesis.

To account for statistical fluctuations, the procedure is repeated multiple times with different random realizations of the observed event rates. The discovery potential is assessed using the discovery Bayes factor, which represents the preference for the signal hypothesis over the background-only hypothesis, using Jeffrey's table. The average Bayes factor over all realizations is reported, indicating the evidence for the signal hypothesis. 

\subsection{Results}

IceCube-Gen2 has the potential to provide decisive evidence for the discovery of most benchmark ultrahigh-energy neutrino flux models within a decade of operation. The results are presented in Fig.~\ref{fig:bayes_factor_hard}, which shows the evolution of the mean discovery Bayes factor with exposure time for flux models 1--12. Our analysis choices are conservative, but alternative choices of background and priors could accelerate the discovery process. Flux models 3--12 are considered with the background of atmospheric muons and the UHE tail of the IceCube high-energy neutrino spectrum, while flux models 1 and 2 only include the background of atmospheric muons. Based on Fig.~\ref{fig:bayes_factor_hard}, the flux models are classified into three categories: models discoverable within 1 year, models discoverable in 1-10 years, and models not discoverable within 20 years. Our results are encouraging and reveal promising prospects for the discovery of an UHE neutrino flux in the first decade of operation of IceCube-Gen2. Several of our benchmark UHE neutrino flux models may even be decisively discovered within 5 years of detector exposure.

In Ref.~\cite{Valera:2022wmu} we explore in detail how our results depend various experimental characteristics, therefore providing a solution to quantify the discovery potential of future neutrino telescopes, and how this depends on their design. Among the more important results in Ref.~\cite{Valera:2022wmu} we find that:
\begin{itemize}
    \item The size of the atmospheric muon background has only a mild impact.
    \item The normalization and the spectral index of the UHE tail of the background IceCube high-energy neutrino flux has a large impact; a softer spectrum yields a smaller background. 
    \item Using an informed prior on the cut-off energy of the background UHE tail of the IceCube high-energy neutrino flux may significantly hasten flux discovery, even if the prior is based on limited knowledge.
\end{itemize}

\section{Flux separation}
\begin{figure*}[t!]
 \centering
 \includegraphics[width=\textwidth]{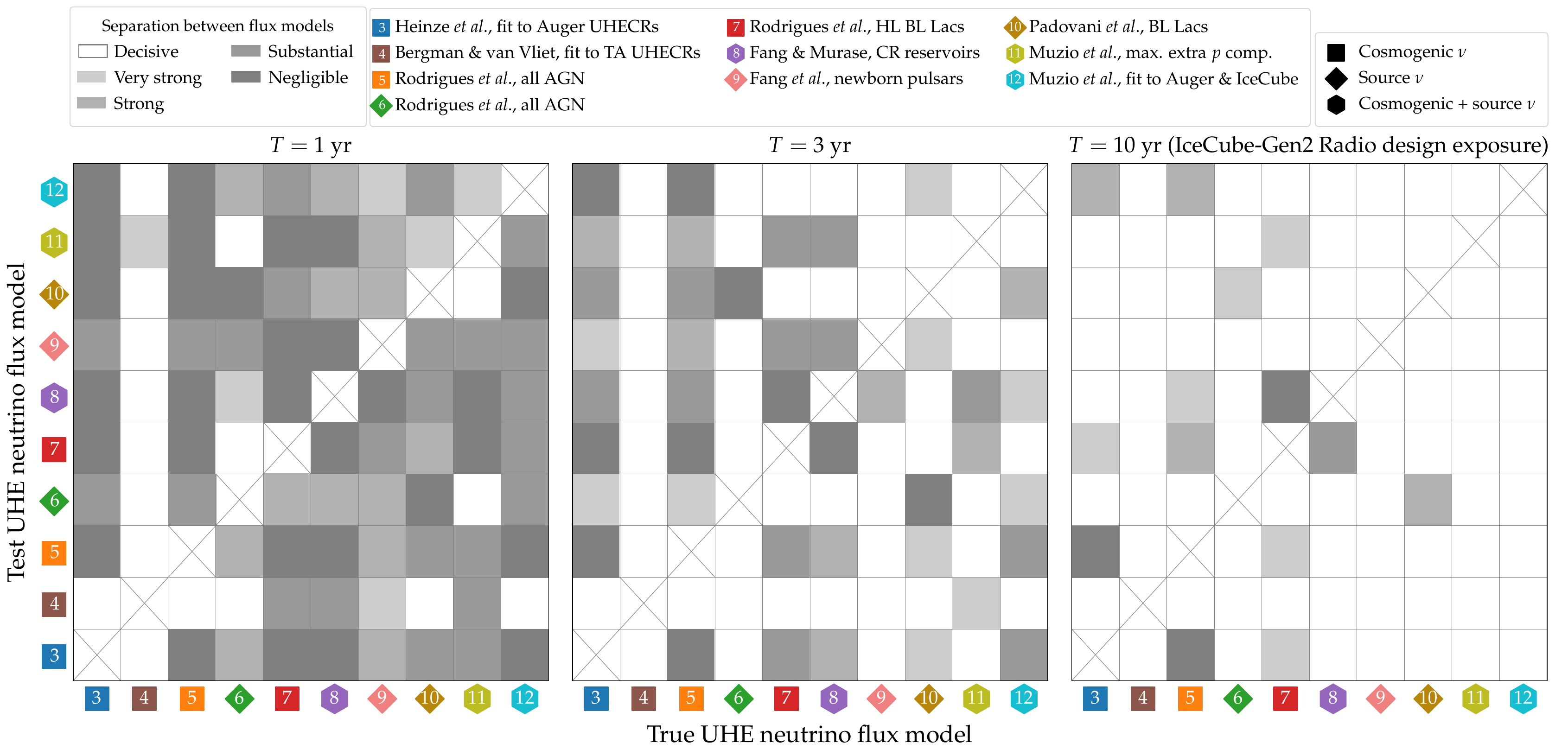}
 \caption{\label{fig:confusion_hard}Confusion matrix showing the experimental separation between true and test UHE neutrino flux models in the radio array of IceCube-Gen2, after an exposure time $T$.  The true flux model determines the observed event rate, and they are contrasted against event-rate predictions from the test models.  The color coding shows the mean model separation Bayes factor, accounting for the background from atmospheric muons and from the UHE tail of the high-energy neutrino spectrum, and interpreted qualitatively using Jeffreys' table. Figure from Ref.~\cite{Valera:2022wmu}.}
\end{figure*}

This section focuses on the distinguishability of two UHE neutrino flux models from each other. The analysis considers the ``true signal hypothesis'', assuming knowledge of the true neutrino flux model, and the ``test signal hypothesis,'' which represents alternative test UHE neutrino flux models. The aim is to forecast how well these hypotheses can be distinguished in the radio array of IceCube-Gen2, specifically for benchmark flux models 3--12. Similar to the previous section, the analysis takes into account background contributions from atmospheric muons and the UHE tail of the IceCube high-energy neutrino flux, statistical fluctuations in the event rate, and uncertainties in analysis parameters.

\subsection{Statistical model}
The statistical analysis for distinguishing between UHE neutrino flux models is based on the methodology used for flux discovery as discussed in the previous section. The likelihood functions under the true and test hypotheses, respectively, are computed using the same procedure of the signal hypothesis for the flux discovery analysis, but with different UHE models for the neutrino signal. The true and test flux models can be any of the benchmark flux models 3--12. The corresponding statistical evidence is computed integrating over the model parameter space the likelihood multiplied by the prior distribution. The model separation Bayes factor is derived as the ratio of the evidences.

\subsection{Results}
Figure~\ref{fig:confusion_hard} presents the confusion matrix, illustrating the ability to distinguish between UHE neutrino flux models after 1, 3, and 10 years of detector exposure time. Each entry in the matrix represents the model separation Bayes factor, interpreted qualitatively using Jeffreys' table. At short exposure times, most flux models cannot be distinguished due to low event rates and poor resolution of energy and angular features. However, with longer exposure times, the event rate increases, leading to better resolution and robustness against fluctuations. As a result, the features of different flux models become more distinct, allowing for clearer differentiation. After 10 years, many flux models that can be discovered also exhibit distinguishable features with strong or decisive evidence. Some exceptions include flux models 7 and 8, which remain easily confused due to similarities in their energy spectra. Models with low event rates and long discovery times, such as models 1, 3, and 5, cannot be distinguished from each other.

%
%
%

\end{document}